\documentclass[]{aa}
\usepackage{graphics}
\begin{document}
\def\trule{\noalign{vskip6pt\hrule\vskip2pt\hrule\vskip6pt}}
\def\brule{\noalign{\vskip6pt\hrule}}
  \thesaurus {03
                 (
                 03.09.7; 
                 03.13.2; 
                 03.20.1; 
                 03.20.4; 
                 05.01.1; 
                 08.09.1; 
                 10.03.1  
                 )}
\title{Analysis of isoplanatic high resolution stellar fields by the StarFinder 
code}
\author{ E. Diolaiti \inst{1} \and O. Bendinelli \inst{1} \and  
D. Bonaccini \inst{2} \and 
L. Close \inst{2} \and  D. Currie \inst{2} \and  G. Parmeggiani \inst{3} }
\offprints{E.Diolaiti\\email:diolaiti@bo.astro.it}
\institute{ Universit\'a di Bologna, Dipartimento di Astronomia, 
via Ranzani 1 - I-40127 Bologna Italy \and
European Southern Observatory, Karl-Schwarzschild Str. 2 - D-85748 
Garching Germany \and
Osservatorio Astronomico di Bologna, via Ranzani 1 - I-40127 Bologna Italy
}
\date{}
\titlerunning{Analysis of isoplanatic high resolution stellar fields by the 
StarFinder code}
\authorrunning{E. Diolaiti et al.}
\maketitle
\begin{abstract}
We describe a new code for the deep analysis of stellar fields,
 designed for Adaptive Optics (AO) Nyquist-sampled images
 with high and low Strehl ratio. 
The Point Spread Function (PSF) is extracted directly from
 the image frame, to take into account the actual structure of 
the instrumental response and the atmospheric effects. The code
 is written in IDL language and organized in the form of 
a self-contained widget-based application, provided with a
 series of tools for data visualization and analysis. A description
 of the method and some applications to AO data
 are presented.
\keywords{instrumentation: adaptive optics - method: data analysis - techniques: 
image processing - techniques: photometric - astrometry - stars: imaging - 
galaxy: center 
}
\end{abstract}
\section{ Introduction}
StarFinder is a code developed within the frame of the ESO PAPAO (Currie et al. 
\cite{currie1}, \cite{currie2})
program for the reduction of Adaptive
Optics data. It was originally designed (Diolaiti et al. \cite{dio1}) to 
analyze AO
images of very crowded fields, like the PUEO frame of the Galactic
Center  shown in this paper, which contains about 1000
detectable stars in a field of view of 13$\times$13 arcsec$^2$. This image is an 
example of an  AO observation: the PSF is Nyquist-sampled and characterized by a
complex shape, showing a sharp peak, one or more fragmented diffraction rings
and an extended irregular halo. Moreover, due to the small field of view, the
imaged region is approximately isoplanatic and the PSF may be considered space
invariant. Under the assumptions of isoplanatism and Nyquist-sampling, 
StarFinder
models the observed stellar field as a superposition of shifted scaled replicas
of the PSF lying on a smooth background due to faint undetected stars, possible
faint diffuse objects and noise.

The procedure derives first a PSF digital template from the brightest isolated
field stars; then a catalogue of presumed objects is formed,
searching for the relative intensity maxima in the image frame.
In the following step the images of the suspected stars are analyzed in order
of decreasing luminosity. In this phase a catalogue including the accepted
objects is formed and a synthetic image of the observed field is constructed,
placing an intensity-scaled PSF template in the position of each identified
star. Each suspected object in the original list is accepted on the basis of
its correlation coefficient with the PSF template; the relative astrometry and
photometry of the source are determined by means of a fit, taking into account
the contribution of the local non-uniform background and of the already
detected stars, known from the synthetic field. As the analysis proceeds,
fainter and fainter sources can be successfully analyzed, discriminating their
central peaks from the secondary bumps of light in the distorted diffraction
rings of the neighboring more luminous already identified stars: in this way
the synthetic field becomes more and more similar to the observed image.
Residual unexplained features of the image may be further
analyzed checking for indications of blended groups, at separations smaller
than the PSF FWHM.

StarFinder should be regarded as a tool for high precision astrometry and
photometry of crowded fields acquired under the above assumptions of accurate
PSF knowledge, isoplanatism and correct sampling. An application of StarFinder 
to HST undersampled data
handled by dithering techniques (Fruchter et al. \cite{fruchter}) has been shown 
in
Aloisi et al. (\cite{aloisi}). In this paper it can be seen that the results 
obtained by our method are comparable to those obtained by DAOPHOT (Stetson 
\cite{stetson}).

As far as the PSF is concerned, the accurate knowledge of its features outside
the central peak is fundamental to perform a deep study of a stellar field,
achieving accurate photometry of faint stars and avoiding either false
detections or star loss (Esslinger \& Edmunds \cite{esslinger}). If the PSF 
template 
cannot be
extracted directly from the field, due to extreme crowding or lack of bright
isolated stars, StarFinder can still be applied using a PSF estimated by means
of other methods, as the reconstruction technique proposed by V$\acute{e}$ran et 
al. (\cite{vera1})
for AO observations or the TINY TIM simulation software for HST 
(Krist \& Hook \cite{krist}).

Much more intriguing and difficult to solve is the case of a field with space
variant PSF, either due to anisoplanatic effects as in AO observations or to
design and control in HST images. In general the analysis of an anisoplanatic
field requires the knowledge of the local PSF. A method to reconstruct the
off-axis PSF in AO imaging has been proposed by Fusco et al. (\cite{fusco}). 
StarFinder,
in its present version, can analyze frames with space invariant PSF or sub-
frames
in which the isoplanicity condition is nearly satisfyied, as will be shown in
Section 4; the extension to the space variant case is in progress, with
preliminary results presented by Diolaiti et al. (\cite {dio2}, \cite{dio3}).

The paper is 
organized as follows: the general features of the algorithm are described
in Sect.~2; Sect.~3 deals with more technical aspects and might be 
skipped on a first reading; the method is validated on simulated and 
experimental data in Sect.~4; details on the IDL code are presented in Sect.~5; 
Sect.~6 
includes our conclusions and future plans.
\section{ Code description}
\subsection{ PSF estimation}
The accuracy of the PSF estimate is primary in StarFinder, since the PSF 
array is used as a template for all the stars in the image to be analyzed. 
The user selects the most suitable stars, which are background subtracted, 
cleaned from the most contaminating sources around, centered with sub-pixel 
accuracy, normalized and superposed with a median operation. The centering 
is performed by an iterative shift of the stellar image in order to cancel 
the sub-pixel offset of its centroid (see Christou \& Bonaccini 
\cite{christou}). 
The median operation, which is performed pixel-by-pixel, is preferred to 
the mean because it is less sensitive to anomalous pixels (outliers) which 
might appear in one or more stellar images among the selected ones. The 
retrieved PSF is post-processed in order to reject any 
residual spurious feature and to smooth the noise in the extended halo. 

The PSF extraction procedure also reconstructs the 
core of saturated stars by replacing the corrupted pixels with the central part 
of the PSF estimate. Accurate positioning is achieved by means of 
a cross-correlation technique, while the scaling factor is determined
with a least squares fit to the wings of the star to 
repair. For a detailed description of the procedure, see Sect.~3.5.
\subsection{Standard analysis of a stellar field }
\begin{figure}
\resizebox{\hsize}{!}{\includegraphics{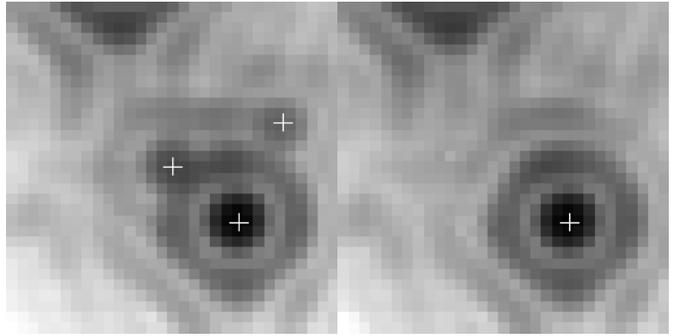}}
\caption{Left: sub-image extracted from a simulated stellar field. The
crosses indicate the objects within the region of interest: a central star (to 
be analyzed in the current step), a brighter source (already known), a fainter 
one (to be examined later) and the PSF feature of a much brighter star, 
represented by the structure in the upper-left part of the sub-image. 
Right: corresponding sub-region 
extracted from the stellar field model, containing one replica of the PSF for
each star detected so far.}
\label{figuno}
\end{figure}

At first we build a list of objects, the presumed stars, which satisfy the 
condition
\begin {equation}
i(x,y)>b(x,y)+t
\end{equation}
where $i(x,y)$ is the observed intensity, $b(x,y)$ the background emission 
and $t$ a fair detection threshold. 
The presumed stars are analyzed one 
by one by decreasing intensity. To illustrate a generic step of the 
analysis, we consider the $(n+1)-th$ object in the list, after the examination 
of 
the first $n$. A small sub-image of fixed size is extracted around the 
object (Fig.~\ref{figuno}, left). This sub-image may contain brighter stars 
formerly analyzed, fainter objects 
neglected in the current step and features of other stars lying outside the 
sub-image. The information on the brighter sources is recorded in a synthetic 
stellar field (Fig.~\ref{figuno}, right), defined as the sum of two terms: 
 a superposition of PSF replicas, one for each star detected up to this point, 
and an estimate of the background, assumed to be non uniform in 
general.
%
The local contribution due to the brighter stars and the background, derived 
from the synthetic field, is subtracted from the sub-image.
If a statistically significant residual remains, it 
is compared to the central core of the PSF by means of a correlation 
check. If the correlation coefficient is higher than a pre-fixed threshold then  
the object of interest is rated similar to the PSF and accepted. The 
accurate determination of its position and relative flux is attained by means of 
a local fit, in which the observed sub-image is 
approximated with the multi-component model described in Sect.~3.6.
 The actual size of the fitting region is comparable to the diameter of the 
first diffraction ring of the PSF. This choice ensures that the information 
represented by the shape of a high-Strehl PSF is included in the fitting process 
to achieve better accuracy and prevents the growth of the number of sources to 
be fitted together. For the central object of our example a single component fit 
is performed and the contribution due to the  brighter stars is considered as a 
fixed additive term. A multi-component fit is performed when 
the star is in a very compact group, at separations comparable to the 
PSF FWHM. 
If the fit is acceptable, the parameters of the new detected star are 
saved and those of the already known sources, 
which have been possibly re-fitted, are upgraded. The new star and an upgrade of 
the re-fitted sources are added to the synthetic field.

\begin{figure}
\resizebox{\hsize}{!}{\includegraphics{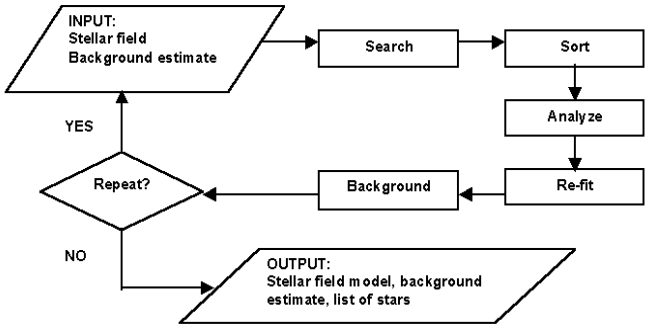}}
\caption{Flow-chart of the algorithm for stars detection and analysis.}
\label{figdue}
\end{figure}

This analysis is performed for each object in the initial list (a flow-chart 
illustrating the operations of StarFinder is in Figure \ref{figdue}). To achieve 
a better astrometric and photometric accuracy, at the end of the examination of  
all the objects, the detected stars are fitted again, this time considering  all 
the known sources. This step may be iterated a pre-fixed 
number of times or until a convergence condition is met.  

At the end of the analysis, it is possible to stop the algorithm or instead 
perform a new search for lost objects removing the detected stars and using
an upgraded background estimate. It should be 
stressed that this image subtraction is just a tool to highlight significant 
residuals. Any further analysis is performed on the 
original frame, in order to take into account the effects arising from the 
superposition of the PSFs of neighboring sources. Generally, after 2-3 
iterations of the main loop, the number of detected stars approaches a stable 
value.
%
\subsection{Crowding and blending effects}
\begin{figure}
\resizebox{\hsize}{!}{\includegraphics{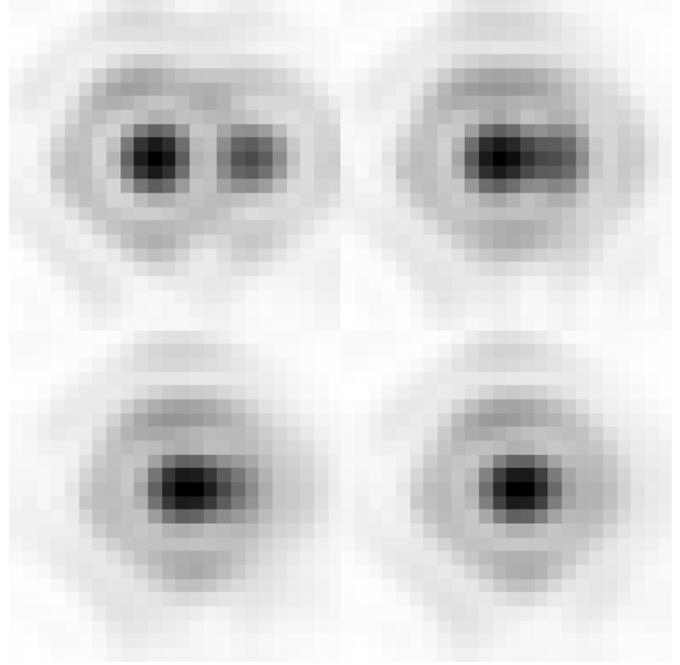}}
\caption{Simulated binary stars at various separations. From left to right, 
top to bottom the separation is 2, 1, 0.75, 0.5 times the PSF 
FWHM. For all the images the flux ratio is 2:1.
}
\label{figtre}
 \end{figure}

A binary star with different separation values (Fig.~\ref{figtre}) has 
been simulated to show how the code works with crowded sources. 
With a  separation of 2 PSF FWHM the two components are well separated and the 
code analyzes them with the standard procedure described in the previous 
sub-section. In the other cases (separation from 1 to 0.5 PSF FWHM) the 
secondary component is not detected as a separate relative maximum and 
it is lost. However, if the separation is not as small as 0.5 FWHM, a further 
iteration of the main loop enables the algorithm to detect the fainter 
component by subtracting the brighter one. This strategy forces the 
two stars to pass the correlation test, the principal component as a 
presumed single object and the secondary in a subsequent iteration of the 
loop.
In a way the iteration of the main loop is a de-blending strategy, 
because it enables the algorithm to detect stars whose intensity peak 
is not directly visible in the observed data.
%
This strategy fails when:
\begin{itemize}
\item the binary is very noisy and the two components have similar flux. In a 
similar situation the principal component may not pass the correlation test, 
preventing also the detection of the secondary in a further iteration. These 
noisy blends may be recovered at the end of the overall analysis.
\item The two components are almost equally bright and have a separation close 
to the lower limit (1/2 PSF FWHM). The residual corresponding to the 
secondary after subtracting the principal component may have a distorted shape 
and might not pass the correlation check.
\end{itemize}

The latter case may be handled by a method based on a thresholding 
technique. The object is cut at a prefixed level, about 20\% below 
the central peak, and transformed to a binary array, setting to 1 all the pixel 
above the threshold and to 0 the pixels below. If the area of the pixels with 
value 1 is more extended than the PSF, the object is considered a blend and 
the secondary star may be detected by subtracting the brighter one; 
then a two-components fit allows accurate astrometry and photometry of the 
two sources. This strategy can be iteratively applied to multiple blends. 
It should be stressed that the area measurement is not reliable when the 
value of the cutting threshold is comparable to the noise level: for this 
reason the de-blending procedure is applied only to objects with a suitable 
signal-to-noise ratio. Moreover the area measurement is reliable when the 
data are adequately sampled. This de-blending procedure is applied at the 
end of the last iteration of the main loop, when all the resolved sources 
have been detected: in this way the probability that a single object may 
appear artificially blurred because of the contamination of still unknown 
sources is largely reduced.

In a normal case, like the simulated field of 
Sect.~4.1, two or three iterations of the main loop find almost all ($\sim$99\%) 
the 
stars that StarFinder may detect. The de-blending procedure described above 
adds $\sim$1\% more stars, without additional false detection. The number of 
lost objects belonging to the first category described above is negligible 
($<<$1\%).

Normally we perform two or three iterations of the main 
loop and apply the de-blending strategy only in very crowded fields.
\section{Technical aspects}
\subsection{Bad pixels}
StarFinder includes an automatic procedure to repair known bad pixels, 
which are replaced with the median of the good data points in a suitable 
neighborhood. The use of this procedure is particularly recommended in the 
extraction of the PSF. Indeed the stellar images, which will be combined to 
form the PSF, are centered with sub-pixel accuracy by means of an interpolation 
technique and the presence of bad pixels may produce unpredictable interpolation 
errors. Even if replaced, the bad pixels are 
excluded from any further computation concerning the correlation coefficient and 
the fitting process.
\subsection{Background estimation}
A reliable estimate of the background is necessary to define
the detection condition and to compute the correlation coefficient of 
the presumed stars with the PSF.

A straightforward technique to estimate the background is to smooth 
the image by median filtering, replacing each pixel 
with the median computed over a suitable neighborhood, of 
size comparable to the characteristic width of a 
stellar image. This method tends to over-estimate the 
background underneath strong peaks.
A more accurate approximation, even in crowded fields, is described 
in Bertin \& Arnouts (\cite{bertin}). The image is partitioned in sub-regions 
arranged in a regular grid and a local estimate is calculated for each 
sub-patch by means of an IDL implementation of the DAOPHOT SKY routine 
(Stetson \cite{stetson}), due to Landsman (\cite{landsman}). This array of 
sky measurements 
is smoothed by median filtering and interpolated onto the same grid of the 
input image.

It should be stressed that the background computation is 
unavoidably affected by the presence of bright sources. In 
general a more accurate estimate can be obtained after the analysis of the 
stellar field, when the most contaminating sources are 
known and can be subtracted.
\subsection{ Noise estimation}
The estimate of the noise is useful to define the detection threshold 
and to compute the formal errors on the retrieved astrometry and photometry.

The overall effect of photon and instrumental noise can be computed if 
the required parameters (detector gain, read-out-noise, dark current, 
etc.) are known. Otherwise an estimate of the mean background noise can 
still be obtained by means of histogram fitting techniques 
(Almoznino et al. \cite{almoznino}, Bijaoui \cite{bijaoui}). 
Assuming that the intensity of the sky radiation is distributed normally
 around a typical value, the histogram of the 
observed intensity levels should be quite similar to a 
gaussian distribution, whose mode and standard deviation 
represent respectively the sky value and the associated 
noise. Actually the contamination due to stellar sources produces a 
high-intensity tail and an artificial broadening of the 
histogram, which prevents an accurate estimate of the background 
noise. This problem can be partially overcome by removing the signal
from the image, leaving only the pixel-to-pixel variations 
associated to pure noise: a reasonable estimate of the signal
 to subtract for this purpose can be obtained by 
smoothing the data with a median filtering technique. 
After this operation the histogram is symmetric around its mode 
and the background noise standard deviation 
can be estimated by means of a gaussian fit to the histogram itself.
\subsection{ Correlation coefficient}
False detections, associated to noise spikes or residual PSF features of
bright stars, are recognized on the basis of their 
low correlation coefficient with the PSF, which represents a template for 
each true star in the image. The correlation coefficient (Gonzalez \& Woods 
\cite{gonzales}) is defined as
\begin{eqnarray}
\lefteqn{c(a,b) = }  \nonumber \\
  &   & \frac{\sum_{x,y}\left[ i\left( x,y\right) -\bar{\imath}\right] \left[ 
p\left(x-a,y-b\right) -\bar{p}\right] }{\sqrt{\sum_{x,y} \left[ i\left( 
x,y\right) -\bar{\imath}\right]^{2}}\sqrt{\sum_{x,y} \left[ p\left( x-a,y-
b\right) -
\bar{p}\right] ^{2}}}
\end{eqnarray}
where $i(x,y)$ and $p(x,y)$ are the object and the PSF respectively,  
$\bar{\imath}$ and $\bar{p}$ are the corresponding mean 
values. Maximizing the correlation coefficient as a function of the offset   
$(a,b)$ yields an objective measure of similarity. 
After maximizing $c(a,b)$  for integral offsets, it is possible to repeat the 
procedure for sub-pixel shifts, improving the 
positioning accuracy.

The correlation coefficient is computed on the core of the PSF: 
typically the central spike of the diffraction pattern is 
considered, out to the first dark ring. A fair correlation threshold must be 
fixed in order to discriminate and reject unlikely detections, without losing 
faint stars contaminated by the background noise; a value of 0.7 or 0.8 is 
acceptable in most cases.

The correlation coefficient represents also an effective tool to 
select the stars with the highest photometric reliability, since generally 
a very high correlation value is associated to resolved single sources.
\subsection{ Saturated stars}%
Saturated stars provide precious information on the PSF halo, so it may be 
useful to include them, appropriately reconstructed, in the PSF extraction 
process. In addition, the repaired 
saturated stars can be recognized by  the star detection algorithm and  
their contribution be taken into account during the analysis of near and fainter 
sources.

The core of a saturated star is replaced with a shifted scaled replica of a 
preliminary estimate of the PSF. The repaired star is defined as
\begin{equation}
i_{rep.}(x,y)=\left\{ \begin{array}{ll}
                  i(x,y)                    &  \mbox{ if $i(x,y)<T$} \\
              f p(x-x_{0},y-y_{0})          &  \mbox{otherwise}
                              \end{array}
                          \right. 
\end{equation}

where $T$ is the upper linearity threshold of the detector. The position of the 
center $(x_{0},y_{0})$ is estimated by means of the 
maximization of the correlation coefficient, which is not sensitive to the 
intensity levels. The scaling factor $f$ is calculated with a least squares fit 
to the wings of the saturated star. Of course the saturated pixels are excluded 
from all the computations. The background is temporarily subtracted in order to 
prevent affecting the repair process.

This procedure has been applied to the brightest star (IRS 7) of the Galactic 
Center 
image shown in Sect.~4.2. This source is not 
saturated in the original data, but it has been artificially corrupted 
by an upper cut at half maximum. The repair procedure 
is able to reconstruct the original peak with an error of $\sim$5\%, 
imposing a positioning accuracy of 1/2 pixel.
\subsection{Fitting procedure}
A sub-image centered on the star of interest is extracted and 
approximated with the model
\begin{eqnarray}
h(x,y) & = & s_{0}(x,y)+\sum_{n=1}^{N_{s}}f_{n}p\left( x-x_{n},y-y_{n}\right) 
\nonumber \\
       &    & +b_{0}+b_{1}x+b_{2}y
\end{eqnarray}
where $s_{0}(x,y)$  is the fixed contribution of known stars outside the sub-
image 
support,  $N_{s}$ is the number of point sources 
within the sub-image, $x_{n}$, $y_{n}$, $f_{n}$  are the position and flux of 
the 
$n - th$   source,   
$p(x,y)$ is the PSF and $b_{0}$, $b_{1}$, $b_{2}$  are the 
coefficients of a slanting plane representing the local background.
It should be stressed that the retrieved astrometry and photometry are referred
to the absolute centering and normalization of the PSF array.
The optimization of the parameters is performed by minimizing the 
least squares error between the data and the model. If 
the noise is known, it is possible to weigh the data by 
their inverse standard deviation and obtain a statistically 
optimal fit (Beck \& Arnold \cite{beck}). In this case, formal error 
estimates on the parameters can be obtained (Bevington 
\& Robinson \cite{bevington}).
The optimization is performed by means of an iterative Newton-Gauss 
technique with linearized Hessian (Beck \& 
Arnold \cite{beck}, Luenberger \cite{luen} ). The Moore-Penrose generalized 
inverse concept (Rust \& Burrus \cite{rust}, Bendinelli et al. 
\cite{bendinelli}, Lorenzutta \cite{lorenzutta}) is applied to invert the 
Hessian matrix at 
every step. The iterations are stopped when the parameters approach a stable 
value.
The major difficulty of a Newton-like method is represented by the 
computation of the model derivatives with respect to 
the parameters, some of which (stellar positions) yield non-linear 
dependence. For this purpose the Fourier shift theorem is applied:
\begin{eqnarray}
\lefteqn{p(x-x_{n},y-y_{n})  = } \nonumber \\
  &   &FT^{-1} \left[FT \left [p(x,y) \right]e^{-i2\pi (ux_{n}+vy_{n})/N}\right]
\end{eqnarray}
where $FT$ represents  the discrete Fourier transform operation, $N$  is the 
sub-image size 
and $u$, $v$  are spatial frequencies. The derivative with 
respect to $x_{n}$  is then
\begin{eqnarray}
\lefteqn{ \frac{\partial p\left( x-x_{n},y-y_{n}\right) }{\partial x_{n}} = } 
\nonumber \\
 & & FT^{-1}\left[ FT\left[ p\left( x,y\right) \right]e^{-i2\pi 
\left(ux_{n}+vy_{n}\right) /N}\left(-i \frac{2\pi u}{N} \right)\right] = 
\nonumber \\
 & & FT^{-1}\left[ \left( -i\frac{2\pi u}{N}\right) FT\left[ p\left(x-x_{n},y-
y_{n}\right) \right] \right] 
\end{eqnarray}
and requires in practice an interpolation of the PSF to compute $p(x-x_{n},y-
y_{n})$ . 
In principle this interpolation-based method can 
only be applied to Nyquist-sampled images and this is 
currently the main limit of the algorithm. A similar technique
has been described by V$\acute{e}$ran \& Rigaut (\cite{vera2}).
\subsection{ Sampling and interpolation}
A band-limited function in 
one dimension is Nyquist-sampled if the step size 
fulfills the condition
\begin{equation}
\Delta x\leq \frac{1}{2}f_{c}
\end{equation}								
where $f_{c}$ is the so-called cut-off frequency of the spectrum. In this case
it is possible to reconstruct 
the original continuous function from a set of equally-spaced 
samples by means of the so-called sinc interpolation 
(Mariotti \cite{mariotti}). For a nearly diffraction-limited PSF, the critical 
sampling condition is generally stated by saying that the 
its FWHM must contain at least two pixels.
Many accurate and efficient interpolation schemes exist to 
perform the PSF shift required by the fitting procedure.
Probably the most simple is a straightforward application 
of the Fourier shift theorem, but one may also use bicubic 
splines or the fast sinc interpolation algorithm described in 
Yaroslavksy (\cite{yaro}). In StarFinder we have applied the cubic 
convolution interpolation method (Park \& Schowengerdt \cite{park}), which 
approaches  very closely the optimal sinc interpolation for Nyquist-sampled
 data; this algorithm is implemented in the IDL function 
INTERPOLATE.

Even though the interpolation of the PSF array is allowed only
on Nyquist-sampled data, 
the cubic convolution technique seems to 
produce acceptable results even with marginally under-sampled 
images. Tests performed on Airy diffraction patterns, 
with a sampling step twice as large as 
the critical sampling step size, indicate that the interpolation-induced 
oscillations amount to a few percent of the 
image peak, as opposed to 10-20\% of other interpolation techniques
 like Fourier shift or bicubic splines. 
\section{ Applications}
\subsection{Simulated field}
\begin{figure}
\resizebox{\hsize}{!}{\includegraphics{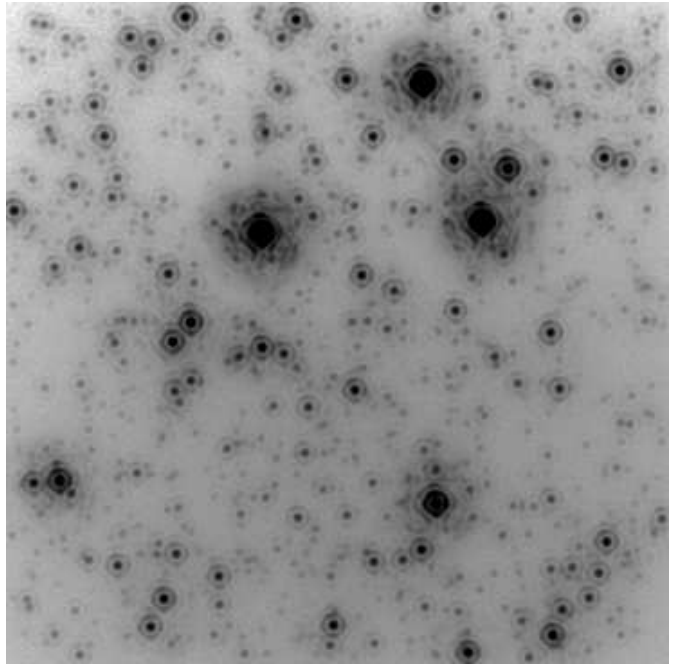}}
\caption{Synthetic field image with 1000 sources. The PSF Strehl ratio is 
$\sim$40\%. The display stretch is logarithmic.}
\label{figquattro}
\end{figure}

\begin{figure}
\resizebox{\hsize}{!}{\includegraphics{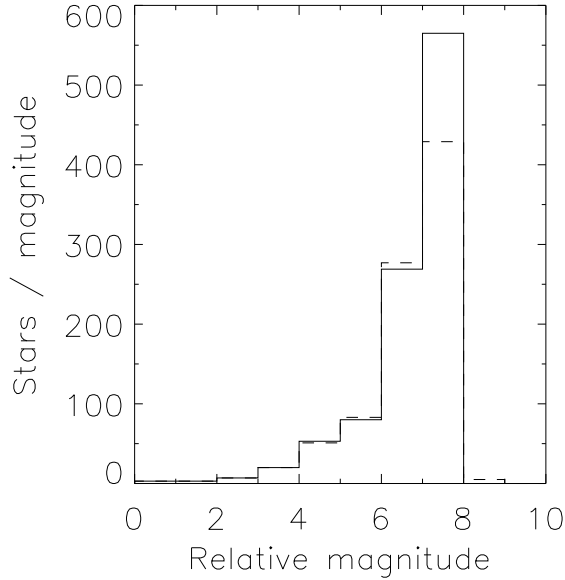}}
\caption{Comparison between the true (continuous line) and the 
estimated luminosity function (dashed line).
}
\label{figsei}
\end{figure}

The code was first applied to a simulated image (Fig.~\ref{figquattro}) 
including 1000 stars, placed randomly in the frame and with a given magnitude 
distribution (Fig.~\ref{figsei}). Each star is a scaled 
copy of a long exposure high-Strehl PSF, obtained with the ADONIS AO system at 
the ESO 3.6~m telescope. The image is 368 $\times$ 368 pixels large 
($13 \arcsec \times 13 \arcsec$) and has a stellar density of 6 stars
arcsec$^{-
2}$.
Photon, readout noise and a background nebulosity, normalized to the 
same flux of the stellar sources, were added to the image. The faintest stars 
have a peak signal-to-noise ratio of~5.
%

We performed a standard reduction of the artificial image using the "default" 
parameters of the method.

\begin{figure}
\resizebox{\hsize}{!}{\includegraphics{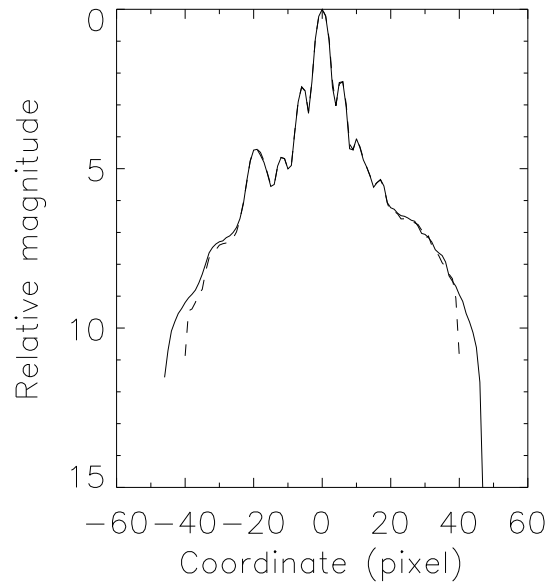}}
\caption{Axial plot of the true PSF (continuos line) and of the retrieved 
PSF (dashed line). 
}
\label{figcinque}
\end{figure}

The PSF was estimated by superposing the images of the four brightest stars in 
the field. The retrieved PSF is very similar to the true one 
(Fig.~\ref{figcinque}).
%

Figure \ref{figsei} shows the good  agreement between the true and the
 observed luminosity function. The 
sole statistically 
significant discrepancy is in the bin from magnitude 7 to 8, 
where $\sim$25\% of the stars were lost; the other small differences 
are due to photometric errors which shift some objects to a 
neighboring magnitude interval. The lost stars are $\sim$10\% of the total 
number of sources and are generally faint: about 90\% have magnitude between 
7 and 8, the rest is in the bin between magnitude 6 and 7. The only bright 
lost star has magnitude $\sim$4.5 and is the secondary component of a very 
close binary, with a separation of just 1/2 pixel.
%
Roughly 70\% of the lost stars are located at a distance $\leq$ 1 PSF FWHM 
from the nearest object, $\sim$20\% are on the first diffraction ring of a 
brighter 
source and only $\sim$10\% are isolated. It should be stressed however that 
$\sim$15\% of the lost stars can be recognized by visual inspection as faint 
objects in the halo of the 
four brightest stars in the field, independently of their separation from the 
nearest source. It is apparent that the blending effect and the contamination by 
the halo of very bright stars do account for the lost stars.
Note that the number of false detections in this simulated field is negligible 
(1 case out of 1000).

\begin{figure}
\resizebox{\hsize}{!}{\includegraphics{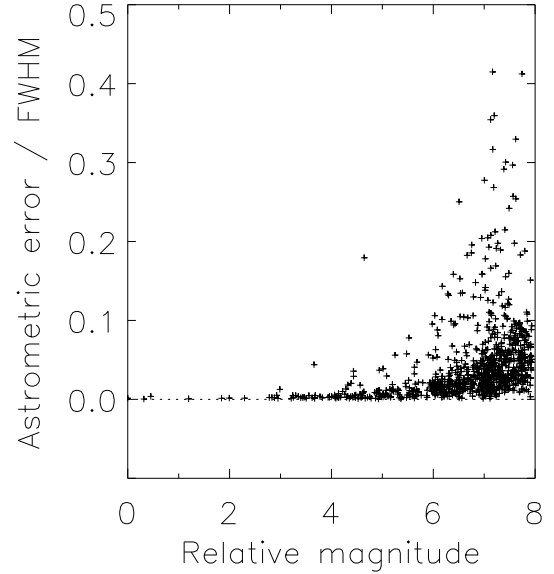}
}
\caption{Plot of astrometric errors vs. relative magnitude of 
detected stars; the errors are quoted in 
FWHM units (1 FWHM $\sim$ 3.6 pixel) and represent the distance 
between the calculated  and the true position. A tolerance of 1 
FWHM has been chosen to find the coincidences between the detected 
stars and their true counterparts. The small errors for the brightest 
stars are due to blending effects.} 
\label{figsette}
\end{figure}

\begin{figure}
\resizebox{\hsize}{!}{\includegraphics{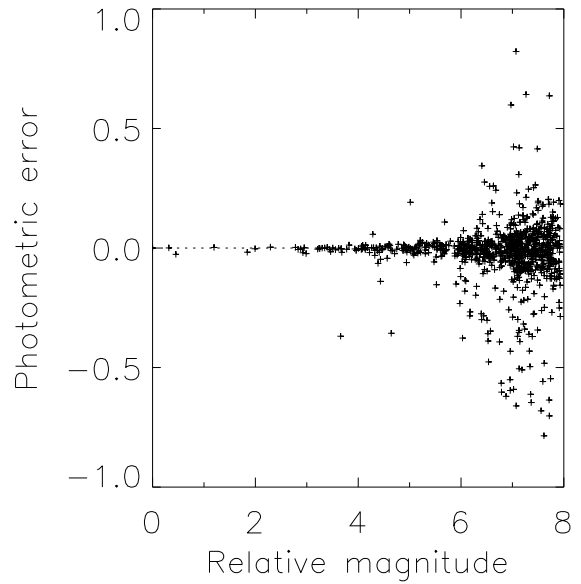}}
\caption{ Plot of photometric errors vs. relative magnitude of 
detected stars. The brightest star in the field has mag=0 
by definition.} 
\label{figotto}
\end{figure}

The plots in Figures \ref{figsette}, \ref{figotto} show the astrometric and 
photometric accuracy of StarFinder. About 80\% of the detected stars 
have both astrometric error smaller than 0.1 PSF FWHM and photometric error 
smaller than 0.1 magnitudes. The stars with less accurate astrometry or 
photometry are generally faint: $\sim$75\% belong to the magnitude interval 
from 7 to 8, $\sim$20\% to the bin between magnitude 6 and 7 and only 
$\sim$5\% are distributed in lower bins of the luminosity function. 
%
These stars have, in $\sim$45\% of the cases, a lost source in their immediate 
neighborhood within the first diffraction ring of the PSF. The others are faint 
stars dispersed in the halo of the brightest sources or in the noisy 
background nebulosity.
%

After discussing the performance of the code with a standard analysis, it is 
interesting to examine how the results are affected by the main parameters of 
the method. Applying the de-blending strategy we detected $\sim$+10\% more 
binaries in the separation range between 1/2 and 1 PSF FWHM, even though the 
overall detection gain is less than 1\% referred to the total number of sources. 
With a higher number of iterations of the main loop we detected $\sim$+30\% 
more binaries in the range from 1 to 2 PSF FWHM. Decreasing the detection 
threshold from 3 to 1 times the noise standard deviation, we found $\sim$+25\% 
more binaries in the range between 1/2 and 1 PSF FWHM, but with 10 faint 
(mag $> 8$) false detections instead of 1. Increasing the threshold on the 
correlation coefficient, from 0.7 to 0.8, we reported no false detection, but 
the number of lost stars increased by about 60\%; the additional lost sources 
were generally fainter than magnitude 7, but not necessarily in crowded groups. 
Lowering the correlation threshold to 0.6 we detected more faint isolated stars 
and binaries, at separations between 1 and 2 PSF FWHM, but with a higher 
probability of false detections (2 instead of 1). Finally the astrometric and 
photometric accuracy approaches a stable level after a few ($\sim$2) re-fitting 
iterations.
\subsection{ Galactic Center}
\begin{figure}
\resizebox{\hsize}{!}{\includegraphics{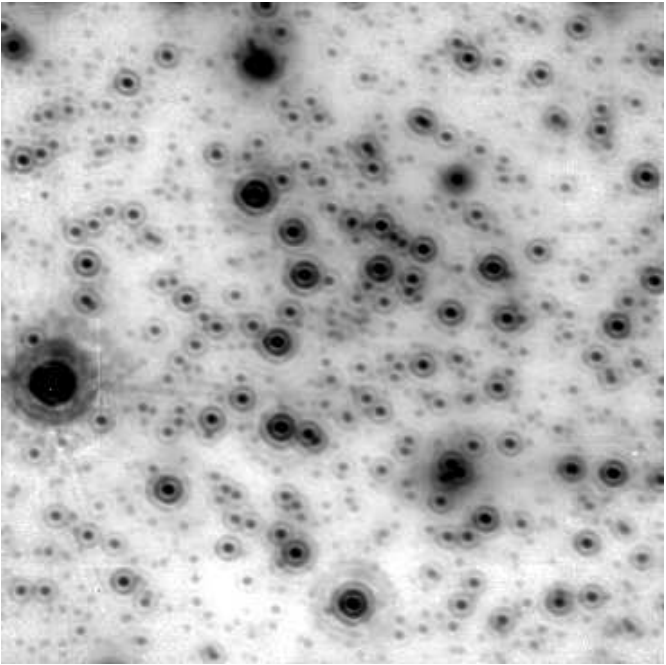}}
\caption{ PUEO image of the Galactic Center. 
North is to the left (at an angle of -100.6\degr from vertical)
and east is -10.6\degr from the vertical.  The display stretch is logarithmic.
\label{fignove}
}
\end{figure}

The code was run on a 15 min exposure time K band (2.2$\mu$m) image of the Galactic 
Center (Fig.~\ref{fignove}),  taken with the PUEO AO system on the 3.6m CFH 
telescope (Rigaut et al. \cite{rigaut}). The Strehl ratio in the image is $\sim$45\%.
The PSF FWHM is $\sim 0.13 \arcsec$,  with a sampling 
of $0.034 \arcsec$ /pixel.
The Adaptive Optics guide star was a $m_{R}$=14.5 star (called star 2 in 
Biretta et al. \cite{biretta})
located about (to the upper 
left) $20 \arcsec$ from the center of the image, out of the field of view of the figure. 
There is therefore a slightly elongation of the PSF towards the 
direction of the guide star. 
However, due 
to the fact that the isoplanatic patch was much larger than the
 $13 \arcsec \times 13 \arcsec$ shown in 
the figure, a space-invarient PSF fits the data very well, as we will show.

%
%
%
%
\begin{figure}
\resizebox{\hsize}{!}{\includegraphics{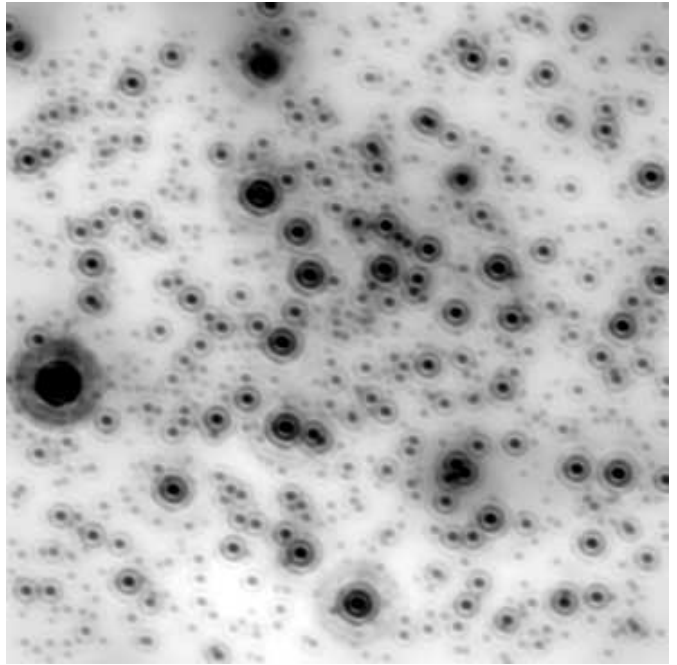}}
\caption{ Reconstructed image, given by the sum of about 1000 detected stars 
and the estimated background. The display stretch is logarithmic.
\label{fignovea}
}
\end{figure} 

A standard analysis was performed, analogous to the 
one described in Sect.~4.1 for the synthetic stellar field. 
About 1000 stars were detected, with a correlation coefficient 
of at least 0.7; the reconstructed image is shown in 
Figure \ref{fignovea}.

\begin{figure}
\resizebox{\hsize}{!}{\includegraphics{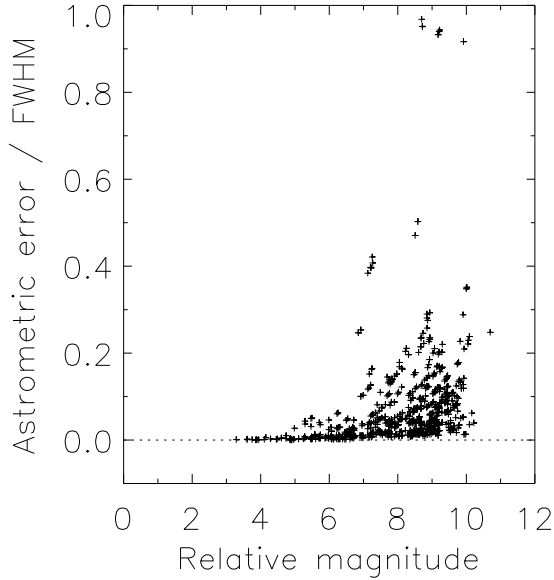}}
\caption{Plot of astrometric errors vs. relative magnitude 
of detected synthetic stars; the errors are quoted in 
FWHM units (1 FWHM $\sim$ 4 pixel) and represent the distance 
between the calculated and the true position.}
\label{figdieci}
\end{figure}

\begin{figure}
\resizebox{\hsize}{!}{\includegraphics{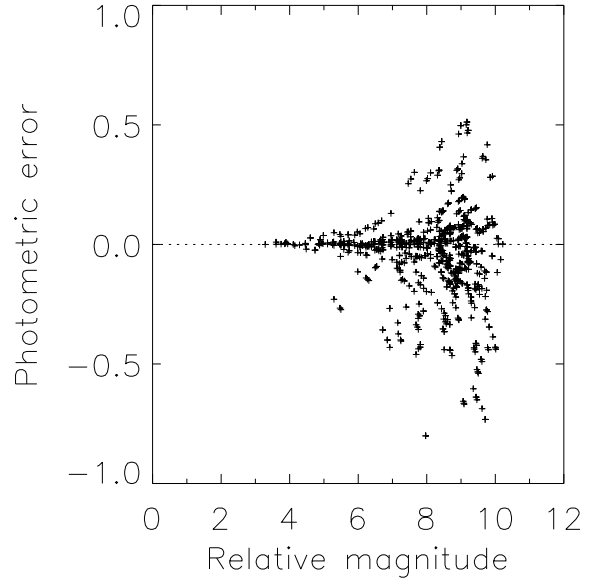}}
\caption{Plot of photometric errors vs. relative magnitude 
of detected synthetic stars. The brightest star in the field has 
mag=0 by definition.}
\label{figundici}
\end{figure}

\begin{figure}
\resizebox{\hsize}{!}{\includegraphics{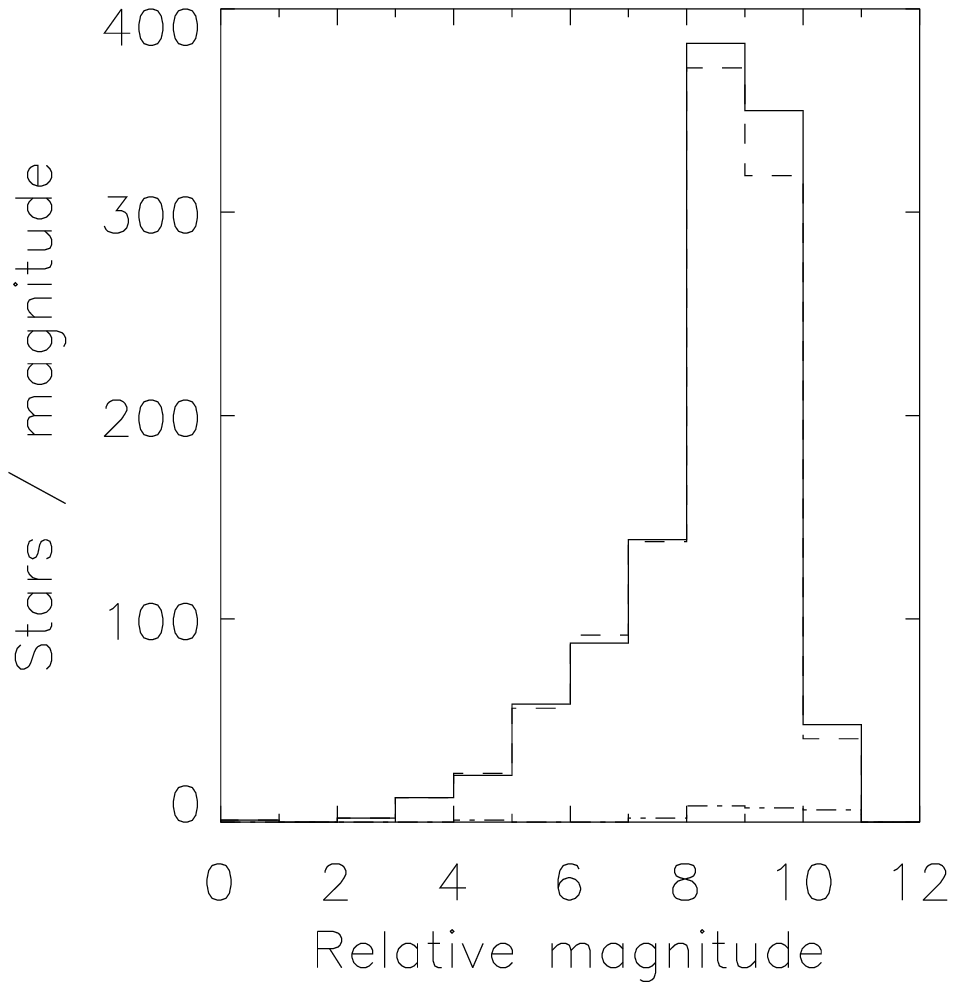}}
\caption{Luminosity function after adding synthetic 
stars (continuous line) compared to the mean luminosity 
function resulting from the analysis of the 10 frames with 
artificial sources (dashed line); the dotted-dashed line indicates 
the mean luminosity function of the false detections.
}
\label{figdodici}
\end{figure}

We evaluated the accuracy of the algorithm by means of 
an experiment with synthetic stars. We created 10 frames adding 
to the original image 10\% of synthetic stars at random positions 
for each magnitude bin of the estimated luminosity function. The 10 frames 
were analyzed separately. As in the simulated 
case, a distance tolerance of 1 PSF FWHM was adopted to find 
coincidences between the detected stars and their true counterparts.
The lists of detected artificial stars 
were merged together and the astrometric and photometric errors 
were computed and plotted as a function of the true magnitude 
(Figs.~\ref{figdieci}, \ref{figundici}). 
%
%
%
The plots show no apparent photometric bias and high astrometric 
and photometric accuracy: the stars brighter than magnitude 5, for instance, 
have a mean astrometric error $<0.5$ mas and a mean absolute photometric error 
$<0.01$ mag.
It should be stressed however that the artificial sources are 
contaminated by the background noise 
present in the observed data and by the photon 
noise due to neighboring stars, but no additional noise was added.
Figure \ref{figdodici} shows a comparison between the mean luminosity function 
retrieved in the 10 experiments and the truth. Assuming an 
expected error for each bin equal to the square root of the corresponding 
number of counts, according to the Poisson 
statistic, the only significant differences occur for magnitudes fainter 
than 9.
%
It is also interesting to consider the magnitude distribution of the false 
detection cases (dashed-dotted line in Figure \ref{figdodici}), i.e. the 
detected stars for which we found no counterpart in the original list, 
within a distance of 1 PSF FWHM. The mean percentage of false detections in the 
10 
experiments is 2\% of the total number of stars. The false detections are almost 
always very faint (mag $>8$); their number is comparable to the square root of 
the 
total counts only in the last magnitude bin, for magnitudes fainter than 10. The 
percentage of false 
detections reported in these experiments seems to confirm the analysis performed 
by 
visual inspection on the stars detected in the original frame.
\section{ The StarFinder code}
The StarFinder code has been provided with a collection
 of auxiliary routines for data visualization and basic image 
processing, in order to allow the user to analyze a stellar field, 
produce an output list of objects and compare different 
lists, e.g. referred to different observations of the same target. 
The input image is supposed to be just calibrated.

The code is entirely written in the IDL language and has been 
tested on Windows and Unix platforms supporting IDL v. 
5.0 or later.
A widget-based graphical user interface has been created. 
The main widget appearing on the computer screen is nothing 
more than an interface to call secondary widget-based 
applications, in order to perform various operations on the image. 
The basic documentation about the code can be found in
the on-line help pages.
IDL users might wish to run interactively the StarFinder routines, 
without the widget facilities: complete documentation 
on each module is available for this purpose. 
\section{ Conclusions and future developments}
The elaboration of real and simulated data seems to prove the
effectiveness of StarFinder in analyzing crowded isoplanatic stellar fields 
characterized by high Strehl ratio PSFs  and
correct sampling, reaching in this case the full utilization of the data 
information content. The code can be applied also to low Strehl data with 
results comparable to those attainable by other methods. Moreover we 
are evaluating its performance on undersampled images.

StarFinder is also reasonably fast: the 
analysis of the Galactic Center image ($368 \times 368$  pixels, $\sim$1000 
stars) 
requires between 5 and 10 minutes on a normal PC (Pentium 
Pro - 64Mb RAM - 350MHz). The graphic interface makes it 
accessible to users unfamiliar with IDL.

In the near future we plan to provide the code with tools to 
handle images with spatially variable PSF and new elements 
which might be helpful in the analysis of a stellar field.
An interesting application, still in progress, aims at cleaning mixed fields 
from
the contamination of foreground stellar images, leaving out true or suspected
diffuse sources. 
\par
\par
The StarFinder package and its technical documentation
are directly available at: {\tt http://www.bo.astro.it}
or can be obtained on 
request to E. Diolaiti. 

\begin{acknowledgements}
Fran\c{c}ois Rigaut is acknowledged for kindly providing the PUEO image of the 
Galactic Center and for supporting the initial development of this method.
The ESO La Silla 3.6m telescope team is acknowledged for the support during 
the technical observations with Adonis, from which we have retrieved 'real' PSFs 
used 
to simulate stellar fields.
This work was partially supported by the ESO Instrumentation Division
Adaptive Optics Dept. 
and  by  the  Italian Ministry for University 
and Research (MURST) under grant Cofin 98-02-32.
\end{acknowledgements}

\end{document}